# Very High Interfacial Thermal Conductance in Fully hBN-Encapsulated MoS$_2$ van der Waals Heterostructure

Fan Ye[1], Qingchang Liu[2], Baoxing Xu[2*], Philip X.-L. Feng[1,3*], Xian Zhang[4*]

[1]Department of Electrical Engineering & Computer Science, Case School of Engineering,
Case Western Reserve University, Cleveland, OH 44106, USA

[2]Department of Mechanical and Aerospace Engineering,
University of Virginia, Charlottesville, VA 22904 USA

[3]Department of Electrical & Computer Engineering, Herbert Wertheim College of Engineering,
University of Florida, Gainesville, FL 32611, USA

[4]Department of Mechanical Engineering,
Stevens Institute of Technology, Hoboken, NJ 07030 USA

*Abstract*

We report experimental and computational studies of thermal transport properties in hexagonal boron nitride (hBN) encapsulated molybdenum disulfide (MoS$_2$) structure using refined optothermal Raman techniques, and reveal very high interfacial thermal conductance between hBN and MoS$_2$. By studying the Raman shift of hBN and MoS$_2$ in suspended and substrate-supported thin films under varying laser power and temperature, we calibrate lateral (in-plane) thermal conductivity of hBN and MoS$_2$ and the vertical interfacial thermal conductance in the hBN/MoS$_2$/hBN heterostructure as well as the interfaces between heterostructure and substrate. Crucially, we have found that interfacial thermal conductance between hBN and encapsulated MoS$_2$ is 74 ± 25 MW/m$^2$K and 72 ± 22 MW/m$^2$K in supported and suspended films, respectively, which are significantly higher than interfacial thermal conductance between MoS$_2$ and other substrates. Molecular dynamics (MD) computations conducted in parallel have shown consistent results. This work provides clear evidence of significantly efficient heat dissipation in hBN/MoS$_2$/hBN heterostructures and sheds light on building novel hBN encapsulated nanoelectronics with efficient thermal management.

**Introduction**

Atomically thin crystalline materials [1,2,3,4,5,6,7] have spurred increasing research interests owing to their attractive electronic [1,2,3,7,8], optical [4,7,9,10,11], mechanical [12], and energy harvesting [13,14] properties for enabling novel two-dimensional (2D) devices that hold promises for future information technologies [13,15,16,17]. The advances in mechanical assembly techniques [3,8,18] and synthesis methods [19] empower researchers to construct van der Waals heterostructures by creating and stacking various 2D layers together. Among several van der Waals heterostructure platforms, hexagonal boron nitride (hBN) encapsulated structures, sandwiching one or several 2D layers within two hBN layers (*e.g.*, hBN/MoS$_2$/hBN), have been demonstrated with significant improvements in device performance [3,8,18,20,21,22]. In comparison with the mainstream Si or SiO$_2$ substrate, hBN, thanks to its strong in-plane ionic bonding of the planar hexagonal lattice structure, offers a flatter surface which screens the dangling bonds and Coulomb scattering effectively [3]. In electronic domain, fully hBN encapsulated MoS$_2$ field-effect transistors (FETs) with graphite electrodes exhibit Hall mobility up to 34,000 cm$^2$/Vs, which is ten-fold enhancement compared with MoS$_2$ FETs on bare SiO$_2$ [8]. A similar electron mobility boost is also observed in hBN encapsulated graphene FETs [18]. In photonic domain, hBN encapsulated graphene devices have been demonstrated as a high-speed electro-optic modulator with a modulation depth of 3.2 dB and a cutoff frequency of 1.2 GHz [21]. Moreover, thanks to hBN's excellent thermal transport properties [23,24,25,26,27,28], electrical insulation, and chemical inertness, hBN encapsulated devices exhibit excellent stability in harsh environments (*e.g.*, high humidity and elevated temperature up to 200°C) over a long term [22].

Though superior electronic performance has been demonstrated in hBN encapsulated structures, thermal properties, including both lateral (in-plane) and interfacial (out-of-plane) thermal transport in hBN encapsulated structures, remain unexplored. Moreover, in the present effort on achieving high-performance 2D devices, low interfacial thermal conductance between 2D materials and substrates, resulting in inefficient heat dissipation, is regarded as one of the main limitations [29]. For example, the interfacial thermal conductance between $MoS_2$ and $SiO_2$ is in the range of 0.4 - 14 $MW/m^2K$ [29,30] and thus the heat dissipation in $MoS_2$ FETs on $SiO_2$ is quite inefficient, which plagues the performance of $MoS_2$ FETs. Besides, in $MoS_2$/hBN heterostructures with hBN only on one side, the interfacial thermal conductance between $MoS_2$ and hBN is around $17.0 \pm 0.4$ $MW/m^2K$ [28], indicating simply stacking $MoS_2$ on top of hBN may not facilitate heat dissipation sufficiently. Therefore, considering the excellent promises of hBN encapsulated 2D devices and inefficient heat dissipation of 2D materials in various device structures, a comprehensive study on thermal transport of hBN encapsulated structure is of great importance.

In this work, lateral and interfacial thermal transport properties in fully hBN encapsulated single-layer $MoS_2$ sandwich structure (hBN/$MoS_2$/hBN) are carefully studied using the refined optothermal Raman technique [30]. We have found the lateral thermal conductivity of top hBN, bottom hBN, and middle $MoS_2$ are $428 \pm 81$ W/mK, $450 \pm 83$ W/mK, and $83 \pm 33$ W/mK, which are similar to values in previous studies [26,30,31]. In opto-thermal Raman technique, Raman laser is able to transmit through and is absorbed by each 2D layer to provide heat source and to obtain temperature information from the reflected Raman signals by each 2D layer. As the laser power is increased, sample is heated, which enables red-shift Raman mode due to thermal softening. Thermal modeling can then be used to extract the thermal conductivity and interfacial thermal conductance from the measured shift rate. Interestingly, the interfacial thermal conductance in the hBN/$MoS_2$/hBN sandwich structure is about $72 \pm 22$ $MW/m^2K$, which is significantly higher than the thermal conductance between $MoS_2$ and other substrates [28,30,32,33,34], suggesting the fully hBN encapsulation platform offers a much efficient heat dissipation than in other structures. Molecular dynamics (MD) calculations of the interfacial thermal conductance in hBN encapsulated structure agree well with the experimental results and provide the theoretical proof. This work not only demonstrates that the hBN fully encapsulated structures provide much efficient heat dissipation for the encapsulated layers, but also sheds light on engineering new-generation nanoelectronics.

**Results and Discussions**

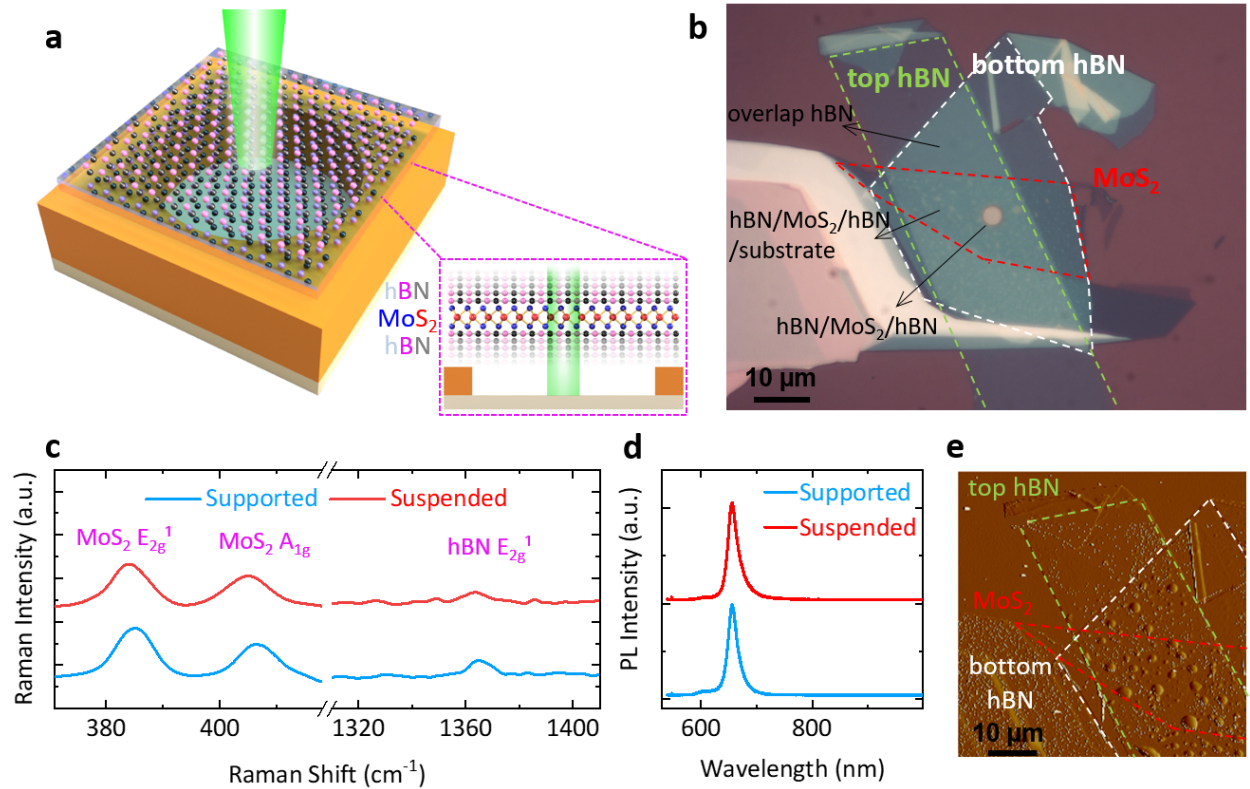

**Figure 1. hBN/MoS$_2$/hBN sample configuration and characterization.** (a) Illustration of a circular drumhead hBN/MoS$_2$/hBN van der Waals heterostructure suspended device and Raman measurement scheme. *Inset*: side view illustration of an hBN/MoS$_2$/hBN device and measurement scheme. (b) An optical microscopy image of an hBN/MoS$_2$/hBN van der Waals heterostructure with top hBN, bottom hBN, overlap hBN, supported hBN/MoS$_2$/hBN/substrate and suspended hBN/MoS$_2$/hBN. (c) Raman, (d) photoluminescence (PL), and (e) atomic force microscopy (AFM) of the hBN/MoS$_2$/hBN van der Waals heterostructure device.

The device structure and measurement scheme are illustrated in Fig. 1a. An hBN/MoS$_2$/hBN van der Waals heterostructure sitting on top of a pre-patterned substrate is heated by a 514 nm laser photothermally. Besides heating up the device, the 514 nm laser is also used to measure the Raman shift which indicates the device temperature and thus calibrate the thermal transport properties. The fabrication of hBN encapsulated MoS$_2$ devices starts from exfoliating hBN and single-layer MoS$_2$ on PDMS to ensure the pristine physical properties. After obtaining the samples on PDMS, bottom hBN, MoS$_2$, and top hBN are stacked one by one on pre-patterned substrates with circular microtrenches (diameter 3 μm and depth 290 nm) using the van der Waals transfer method [35]. A microscope image of an hBN/MoS$_2$/hBN device with 3μm diameter hole is shown in Fig. 1b. After the device fabrication, Raman (Fig. 1c) and photoluminescence (PL) (Fig. 1d) measurements are conducted to characterize the device. From the Raman spectroscopy, two MoS$_2$ characteristic peaks and one hBN characteristic peak are observed in both suspended and supported regions, which confirms the quality of hBN/MoS$_2$/hBN heterostructure in both regions. As there is a clear direct bandgap peak but no indirect bandgap peak in PL, it is confirmed that the MoS$_2$ is one-atomic-layer. Based on atomic force microscopy (AFM) measurements (Fig. 1e), the thicknesses of top hBN and bottom hBN are 10.6 nm and 16.8 nm respectively.

The temperature dependent Raman shift for both MoS$_2$ and hBN was investigated. By gradually increasing the substrate's temperature from 295 K to 495 K, Raman shift of MoS$_2$ and hBN in both supported regions and suspended regions are measured. Fig. 2a presents the schematic of the optothermal Raman technique to measure the lateral thermal conductivity and interfacial thermal conductance. Raman laser is transmitting through each layer of the heterostructure and the reflected spectra include temperature information of each layer. Fig. 2b, c presents an example of the temperature-dependent Raman spectra of supported MoS$_2$ and hBN.

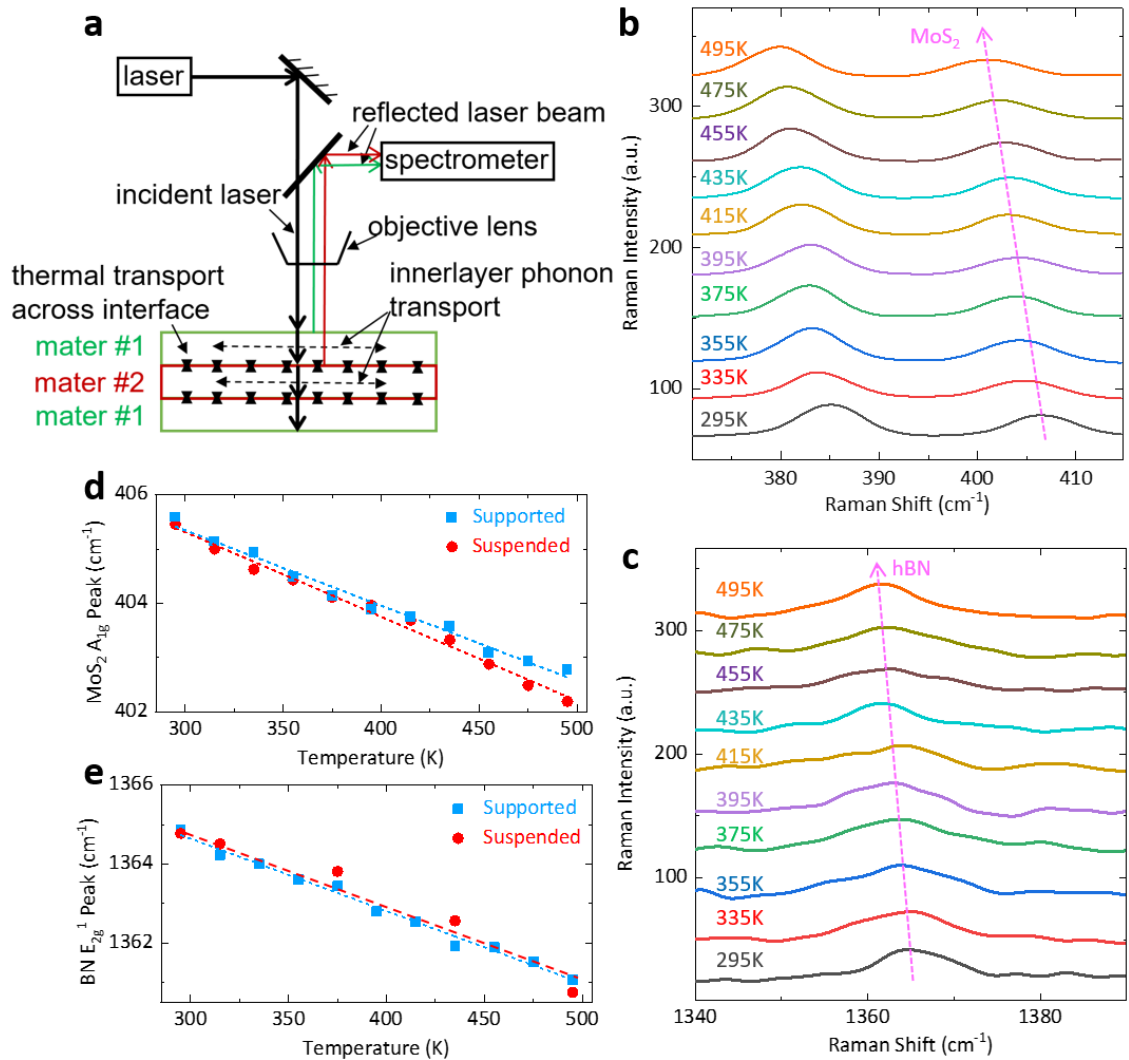

**Figure 2. Schematic of optothermal Raman technique and temperature dependent Raman shifts of hBN/MoS$_2$/hBN heterostructure.** **(a)** Schematic of the optothermal Raman technique. Raman spectroscopy curves of **(b)** MoS$_2$ and **(c)** hBN as temperature increases from 295 K to 495 K. The evolution and fitting of **(d)** MoS$_2$ A$_{1g}$ and **(e)** hBN E$_{2g}^1$ Raman peak position.

The Raman peaks of MoS$_2$ and hBN in both supported and suspended regions show red shift as temperature rises due to the thermally driven bond softening, as previously observed for graphene, MoS$_2$ and MoSe$_2$ [30, 36]. By fitting Raman peaks, the temperature dependence of MoS$_2$ A$_{1g}$ peak shift and hBN E$_{2g}^1$ peak shift is shown in Fig. 2d and Fig. 2e respectively. It is observed that MoS$_2$ A$_{1g}$ peak in suspended region exhibits higher temperature dependence than the supported region, which is attributed to the interaction between MoS2 and substrate as MoS2 $A_{1g}$ peak is out-of-plane vibration mode. In contrast, Raman peak of hBN in suspended region and supported region show similar temperature dependence. By fitting the temperature dependent Raman peak position (Fig. 2c, d) using linear functions, the first-order temperature coefficients (slopes) for hBN and MoS2 in suspended and supported are determined (Table 1).

After obtaining the first-order temperature coefficients for MoS2 and hBN, thermal transport properties of hBN/MoS2/hBN are investigated through photothermal heating the van der Waals heterostructure in both suspended and supported regions. The power adsorbance of hBN and MoS2 to 514 nm laser is calculated and summarized in Table 1. Fig. 3a-d show the Raman shift versus the absorbed laser power of supported MoS2, suspended MoS2, and supported and suspended hBN with top and bottom layers measured simultaneously. By gradually increasing the laser power, Raman peaks of both MoS$_2$ and hBN show red shift as laser power increases, indicating the temperature increase of each layer of the van der Waals heterostructure. For MoS$_2$ and hBN in both suspended and supported regions, Raman peak position exhibits a larger red shift from the 100× objective (spot size 0.19 μm) than the 20× objective (spot size 0.41 μm), which is attributed to the higher power absorbed when a larger spot size is applied. Based on the extracted power dependent Raman peak positions, first-order power coefficient is calculated and summarized in Table 1.

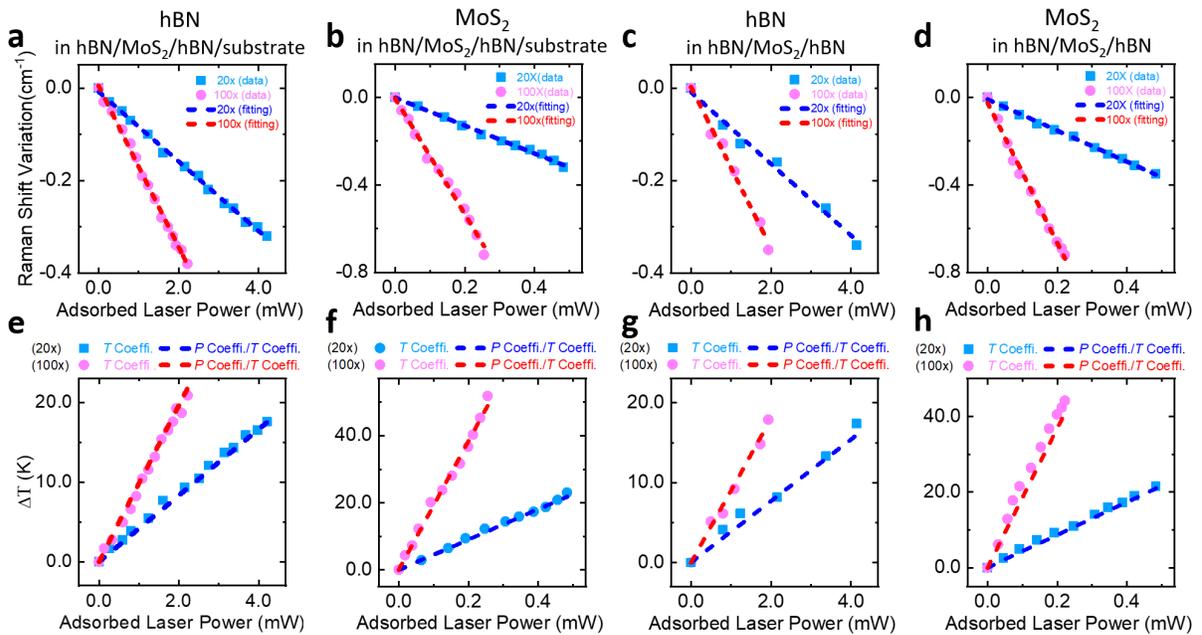

**Figure 3. Power dependent Raman peak shift measured using different laser spot sizes** of **(a)** hBN in supported hBN/MoS2/hBN structure, **(b)** MoS2 in supported hBN/MoS2/hBN structure, **(c)** hBN in suspended hBN/MoS2/hBN structure, and **(d)** MoS2 in suspended hBN/MoS2/hBN structure, **and temperature variation** of these four samples **(e)**, **(f)**, **(g)**, **(h)** induced by photothermal heating corresponding to **(a)**, **(b)**, **(c)**, and **(d)** respectively.

**Table 1**: First-order temperature coefficients, absorbance, and power coefficients of hBN and $MoS_2$ in different structures.

| Interface | Sample location | Temperature coefficient (cm$^{-1}$/K) | Thickness (nm) | Absorbance (%) | Power coefficient (cm$^{-1}$/μW) | |
|---|---|---|---|---|---|---|
| | | | | | 0.19 μm spot | 0.41 μm spot |
| hBN/MoS$_2$/hBN | Supported | MoS$_2$: -0.0139 ± 0.0024 | 0.65 | 3.8 ± 0.1 | -0.1021 ± 0.0081 | -0.0241 ± 0.0017 |
| | | hBN: -0.0183 ± 0.0023 | 27.4 | 33.4 ± 0.2 | -0.0600 ± 0.0060 | -0.0254 ± 0.0023 |
| | Suspended | MoS$_2$: -0.0163 ± 0.0028 | 0.65 | 3.8 ± 0.1 | -0.1150 ± 0.0123 | -0.0271 ± 0.0030 |
| | | hBN: -0.0196 ± 0.0022 | 27.4 | 33.4 ± 0.2 | -0.0588 ± 0.0047 | -0.0251 ± 0.0018 |
| hBN/SiO$_2$ | Top hBN only | hBN: -0.0229 ± 0.0023 | 10.6 | 16.7 ± 0.2 | -0.0857 ± 0.0141 | -0.0338 ± 0.0159 |
| | Bottom hBN only | hBN: -0.0178 ± 0.0030 | 16.8 | 25.0 ± 0.2 | -0.0739 ± 0.0179 | -0.0325 ± 0.0110 |
| | Overlap hBN only | hBN: -0.0232 ± 0.0023 | 27.4 | 37.2 ± 0.2 | -0.1179 ± 0.0273 | -0.0325 ± 0.0171 |

For each objective, the spot size was obtained by scanning across a sharp flake edge, and fitting the measured Raman peak intensity corresponding to the position using a Gaussian error function. The obtained spot size of 100× objective is 0.19 μm, and that of the 20× objective is 0.41 μm.

To determine the absolute power absorbed for heat flow analysis, hBN membranes were separately exfoliated onto transparent quartz substrates, and their measured absorption spectra were used to determine the frequency-dependent complex dielectric function. The dielectric functions of sample, along with that of the substrate material, were then used to calculate the absorption coefficient at 514 nm using the standard transfer matrix method. The absorption coefficient α of hBN can be determined by the measured optical transmittance $T_e$ from the following equation [37]:

$$T_e = \frac{(1-R)^2 exp(-\alpha d)}{1-R^2 exp(-2\alpha d)} \tag{1}$$

where $d$ is the sample thickness, and in the case of normal incidence, $R$ can be expressed simply as a function of the refractive index $n$, $R = (n-1/n+1)^2$, with $n = 1.62$ at 514 nm wavelength. The obtained α is 3730 cm$^{-1}$. Including the thickness $d$, the value of optical absorbance of top hBN, bottom hBN, and overlap hBN are 16.7 ± 0.2 %, 25.0 ± 0.2 %, and 37.2 ± 0.2%, and that of one-atomic-layer $MoS_2$ is 5.2 ± 0.1 % at 514 nm wavelength. For the hBN/$MoS_2$/hBN heterostructure, the transmittance of top hBN is 74.2 % and that of $MoS_2$ is 89.87 %, so the absolute absorbance of $MoS_2$ in the hBN/$MoS_2$/hBN heterostructure is 3.8 ± 0.1 %, and that of the top and bottom hBN considered together is 33.4 ± 0.2 % (Table 1). Due to the low quantum yield of as-exfoliated $MoS_2$ and hBN, the absorbed energy emitting as photons is ignored.

The estimated temperature variation induced by photothermal heating and power absorbed by each 2D layer allow us to calculate the thermal conductivity and interfacial thermal conductance in the hBN/$MoS_2$/hBN van der Waals heterostructure. Based on the Gaussian profile of laser and spot radius of $r_0$, the volumetric heating power density $q'''(r)$ of each layer in the heterostructure is presented as:

$$q'''(r) = P \cdot \frac{1}{t} \cdot \frac{1}{\pi r_0^2} \exp\left(-\frac{r^2}{r_0^2}\right) \tag{2}$$

where $P$ is absorbed laser power, $t$ is sample thickness, $r$ is the distance to laser spot center which varies from 0 to ∞. Considering the interfacial thermal conductance between $MoS_2$ and hBN that's encapsulating it, the temperature distribution *T(r)* in the target layer (MoS₂) of the van der Waals heterostructure is represented using a cylindrical coordinate as:

$$\frac{1}{r}\frac{d}{dr}\left(r\frac{dT(r)}{dr}\right) - \frac{g}{\kappa t}(T(r) - T_a(r)) + \frac{q'''(r)}{\kappa} = 0 \tag{3}$$

where $\kappa$ is thermal conductivity, $g$ is interfacial thermal conductance at the interface between the two material layers with temperatures *T(r)* and *Tₐ(r)* respectively. The boundary conditions of the photothermal heating are described as $\frac{dT}{dr}\big|_{r=0} = 0$ and $T(r \to \infty) = 298\ K$. The measured temperature $T_m$ using Raman shift is the weighted average of the local temperature distribution *T(r)*:

$$T_m = \frac{\int_0^\infty T(r)\exp(-\frac{r^2}{r_0^2})rdr}{\int_0^\infty \exp(-\frac{r^2}{r_0^2})rdr} \tag{4}$$

The measured temperature $T_m$ and absorbed laser power $P$ of each layer has a relation of thermal resistance $R_m$ (K/W) as a ratio between them, which is obtained by dividing power dependent Raman shift (cm⁻¹/W) over temperature dependent Raman shift (cm⁻¹/K). The temperature variations of MoS₂ and hBN as a relation to the absorbed laser power are shown in Fig. 3e-h.

To extract thermal transport properties including lateral thermal conductivity $\kappa$ and interfacial thermal conductance $g$, we apply the same mathematic methods in ref. 30 and ref. 38. In brief, by solving Equations 2–4, we are able to determine a thermal resistance $R_m$ that is a function of intralayer thermal conductivity $\kappa$ and interfacial thermal conductance $g$. Since both MoS₂ and hBN have temperature increase with laser heating, the ratio between the temperature increase of them is used for obtaining the interfacial thermal conductance $g$ between MoS$_2$ and the hBN membranes encapsulating it. Using the first-order power coefficient (cm$^{-1}$/W) and temperature coefficient (cm$^{-1}$/K) $R_m$ (K/W) with 0.19 μm spot size (100×) and 0.41 μm (20×), the lateral thermal conductivity $\kappa$ and interfacial thermal conductance $g$ are calculated and summarized in Table 2. We obtained the lateral thermal conductivities of MoS$_2$ in the hBN/MoS$_2$/hBN encapsulated structure in both supported and suspended regions, bottom hBN alone, top hBN alone, and the overlap hBN of bottom and top hBN samples. The lateral thermal conductivities are 83 ± 33 W/mK (for MoS$_2$), 450 ± 83 W/mK (for bottom hBN), 428 ± 81 W/mK (for top hBN), and 420 ± 74 W/mK (for overlap hBN). The interfacial thermal conductance is 74 ± 25 MW/m$^2$K (between MoS$_2$ and hBN in the encapsulated structure), 69 ± 23 MW/m$^2$K (between top hBN and substrate), 72 ± 22 MW/m$^2$K (between bottom hBN and substrate), and 66 ± 21MW/m$^2$K (between overlap hBN and substrate). Based on the calculation, the thermal conductance of the hBN/SiO$_2$ interface is similar to the previously reported values [39]. In contrast, hBN/MoS$_2$/hBN interface in the encapsulated structure exhibits a much higher interfacial thermal conductance than that in MoS$_2$/SiO$_2$ interface [29,30], suggesting the hBN encapsulation platform provides much efficient heat dissipation for MoS$_2$. The thermal conductivity and interfacial thermal conductance in suspended region are calculated iteratively with thermal transport parameters obtained from the measurements using Eq. 2-4. The lateral thermal conductivity of MoS$_2$ in encapsulated structure in suspended region is 85 W/mK and the interfacial thermal conductance of the hBN/MoS$_2$/hBN interface is 72 ± 22 MW/m$^2$K, which are similar to the values in supported region on the same sample. Here the lateral thermal conductivity of single-layer MoS$_2$ in the hBN encapsulated structure is comparable to that of the suspended MoS$_2$, which is significantly higher than the supported MoS$_2$ on the amorphous substrate (SiO$_2$, Au, etc.) [29,30].

This is explained by the relatively small lattice mismatch and the ultra-clean surface in the van der Waals interface between MoS₂ and hBN [40].

**Table 2**: Lateral thermal conductivities and interfacial thermal conductance in hBN/MoS₂/hBN van der Waals heterostructure device, and the hBN on substrate alone.

| Interface | Sample location | Thermal conductivity (W/mK) | Interfacial thermal conductance (MW/m²K) |
|---|---|---|---|
| hBN/MoS₂/hBN | Supported | MoS₂: 83 ± 33 | 74 ± 25 |
|  | Suspended | MoS₂: 85 ± 36 | 72 ± 22 |
| hBN/SiO₂ | Top hBN only | hBN: 428 ± 81 | 69 ± 23 |
|  | Bottom hBN only | hBN: 450 ± 83 | 72 ± 26 |
|  | Overlap hBN only | hBN: 420 ± 74 | 66 ± 21 |

To gain further insights into the heat transport in hBN encapsulated MoS₂ structure, molecular dynamics (MD) calculation is performed. The device configuration for MD simulation is shown in Fig. 4a. One-atomic-layer MoS₂ sandwiched by two hBN membranes is supported on substrate and a supported fully hBN-encapsulated MoS₂ (hBN/MoS₂/hBN/substrate) structure is therefore constructed. Similar procedures are applied to build the suspended hBN/MoS₂/hBN, hBN/MoS₂, and hBN/MoS₂/Si structures. The intralayer bonded interaction in hBN and MoS₂ is described by the many-body tersoff [41] and Stillinger-weber potential [42], respectively, while the substrate is described by the edip potential. The non-bonded interlayer interaction among hBN, MoS₂ and substrate are described by the 12-6 Lennard-Jones (L-J) potential, $E_{vdW} = 4\epsilon[(\sigma/r)^{12} - (\sigma/r)^6]$, where $\epsilon$ is the potential well depth, $r$ is the distance between different atoms, and $\sigma$ is the parameter associated with the equilibrium distance [43]. The periodic boundary conditions are applied in the x and y directions to eliminate the size effect and free boundary condition was applied in the z-direction to exclude the influence of upper and bottom boundaries. The transient heating technique is employed in the simulations, which has been widely used in the exploration of interfacial thermal conductance of 2D heterostructures in both experiments [28,44] and simulations [43,45]. This method is conducted as follows (Fig. 4b). The system is firstly

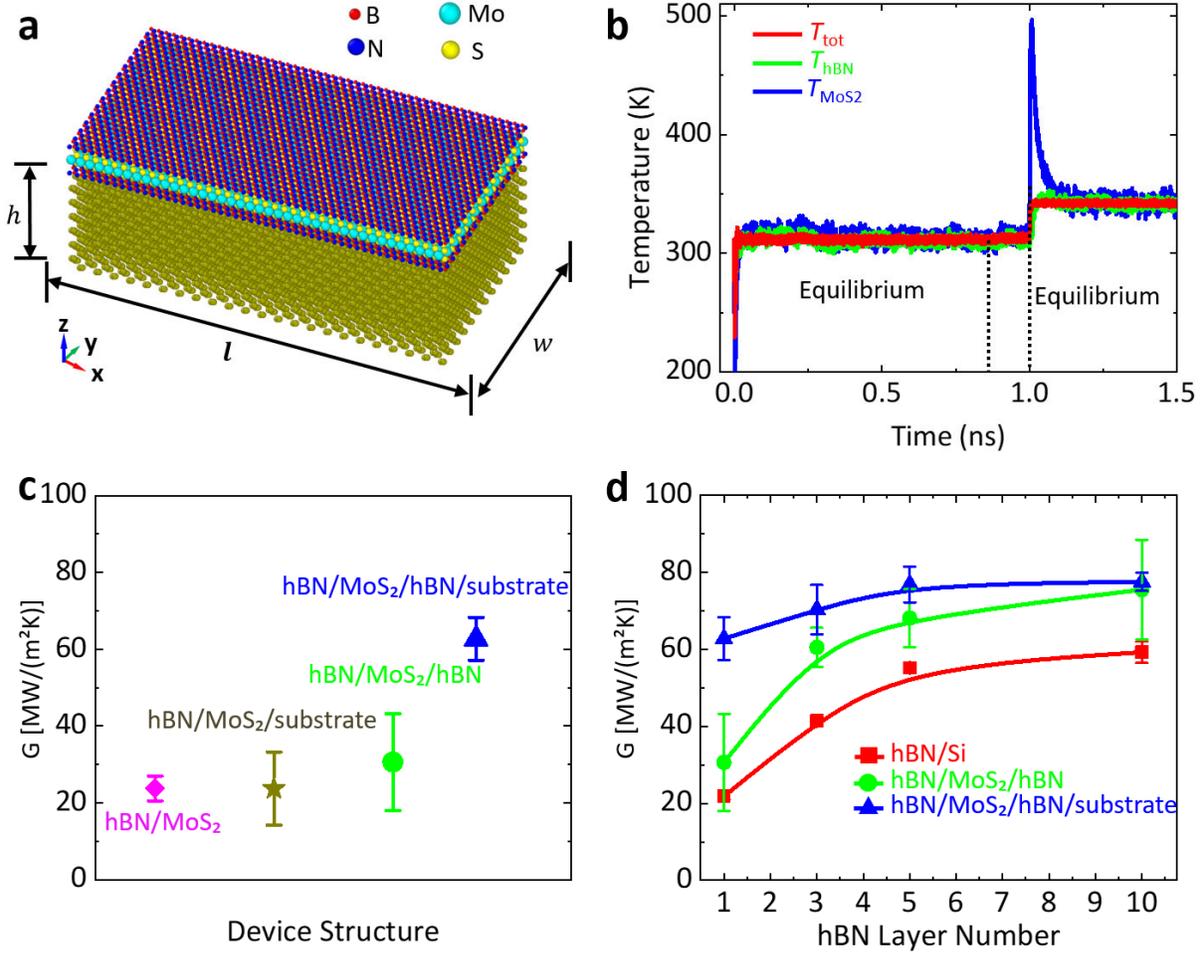

**Figure 4. Molecular dynamics (MD) calculation of interfacial transport properties in MoS$_2$/hBN interfaces. (a)** Illustration of molecular dynamic calculation in the supported hBN/MoS$_2$/hBN/substrate. **(b)** Temperature evolution of MoS$_2$ (heat source, $T_{MoS2}$), hBN (buffer layers, $T_{BN}$) and the system ($T_{tot}$) after the laser excitation. **(c)** Comparison of the interfacial thermal conductance between different structures: suspended hBN/MoS$_2$, supported hBN/MoS$_2$/substrate, suspended hBN/MoS$_2$/hBN, and supported hBN/MoS$_2$/hBN/substrate. **(d)** Variation of interfacial thermal conductance with layer number of hBN for different device structures.

equilibrated under NVT ensemble with a nose-hoover thermostat for 1 ns to make sure the system has reached its initial equilibrium temperature. After that, the NVE ensemble is employed and the MoS$_2$ membrane is exposed to an ultrafast heat impulse, which increases the temperature of MoS$_2$ membrane dramatically in a short time (~10 ps). Since the MoS$_2$ is sandwiched by hBN membranes on both sides, the only path to dissipate thermal energy is from the MoS$_2$ to hBN. As a result, the

-13-

temperature of MoS2 ($T_{MoS2}$) decreases while the temperature of hBN ($T_{hBN}$) increases in the subsequent relaxation process. As the relaxation process continues, the temperature difference between MoS2 and hBN decreases and the system reaches a new equilibrium state after a sufficiently long time (~200 ps). The thermal conductance at this final equilibrium temperature is described as follows. The total energy of MoS2 ($E_{tot}$) is impacted by thermal dissipation. The evolution of energy of $E_t$ is related to the temperature difference between MoS2 and hBN ($T_{MoS_2}$-$T_{hBN}$). The interfacial thermal conductance $g$ can be calculated via $\frac{\partial E}{\partial t} = AG(T_{MoS_2} - T_{hBN})$, where $A$ is the interfacial area and $t$ is time.

Fig. 4c shows the interfacial thermal conductance between hBN and MoS2 for different structures of the system at $T_{tot}$ = 341 K. For suspended hBN/MoS2 structure, the estimated thermal conductance is $g$ = 23.8 MW/m²K, which is very close to that in supported area $g$ = 23.7 MW/m²K. In contrast, the interfacial thermal conductance in suspended hBN/MoS2 is $g$ = 30.6 MW/m²K, indicating that the encapsulated structure promotes the heat dissipation for MoS2. The hBN encapsulated MoS2 heterostructure in supported area – hBN/MoS2/hBN/substrate exhibits interfacial thermal conductance $g$ = 62.7 MW/m²K, which is 160% higher than that in hBN/MoS2/substrate structure and higher than that in suspended hBN/MoS2/hBN. The extra high conductance is ascribed to the synergistic effect of the substrate and the encapsulated structure, where bottom hBN serves as a bridging layer to transport the thermal energy from MoS2 to silicon. However, our experimental results indicate that hBN encapsulated MoS2 in suspended area (hBN/MoS2/hBN/substrate) and supported area (hBN/MoS2/hBN) exhibit very close thermal conductance values (74 MW/m²K and 72 MW/m²K respectively). Considering there are multiple layers (30~50) of hBN in the experiments, it is reasonable to assume thicker hBN exhibits a higher thermal capacity, and as a result, the thermal conductance between hBN and MoS2 become close between the suspended and supported structures. To validate this assumption, a series of simulations with different numbers of hBN layers is performed to probe the interfacial thermal conductance between hBN and MoS$_2$. Fig. 4d presents the variation of thermal conductance with the layer number of hBN in different device structure at $T_{tot}$ = 341K. For the suspended hBN encapsulated MoS$_2$ (hBN/MoS$_2$/hBN) structure, interfacial thermal conductance increases with layer number significantly and reaches 75.5 MW/m$^2$K when the layer number of bottom and top hBN approaches ten. For supported hBN encapsulated MoS$_2$ structure (hBN/MoS$_2$/hBN/substrate), interfacial thermal conductance also increases with hBN layer number enhances, though with a much slower rate compared with suspended structure, and reaches 77.5 MW/m$^2$K when hBN layer number approaches 10. The calculated interfacial thermal conductance is close to experimental results in both suspended structures and supported areas, which validates our experimental results. It is worth noting interfacial thermal conductance between suspended and supported structures become increasingly close with hBN becoming thicker, demonstrating the previous assumption that thicker hBN layer membranes provide rise to a higher thermal capacity.

Finally, we benchmark the results by comparing the interfacial thermal conductance in hBN encapsulated structure and other structures within a temperature range between 295K to 450K [28,30,32,33,34] (Fig. 5). Apparently, the interfacial thermal conductance in hBN/MoS$_2$/hBN van der Waals heterostructure in the vertical direction is significantly higher than MoS$_2$ with other materials. This ultra-high interfacial thermal conductance is mainly attributed to the relatively small lattice mismatch between hBN and MoS$_2$ [46], the high thermal conductivity of hBN, and the excellent encapsulation and contact condition in the hBN/MoS$_2$/hBN van der Waals heterostructure. It is also worth noting that the interfacial thermal conductance in hBN encapsulated MoS$_2$ is higher than the MoS$_2$ on hBN non-encapsulated structure [28]. This could be explained by the fact that the encapsulated structure allows MoS$_2$ to dissipate heat through both top surface and bottom surface, while in the MoS$_2$ on hBN structure, the heat in MoS$_2$ could only dissipate to hBN through the bottom surface. Moreover, the top and down hBN provides a compact and excellent contact conditions in the hBN/MoS$_2$/hBN sandwich structure which ensures a high interfacial thermal conductance.

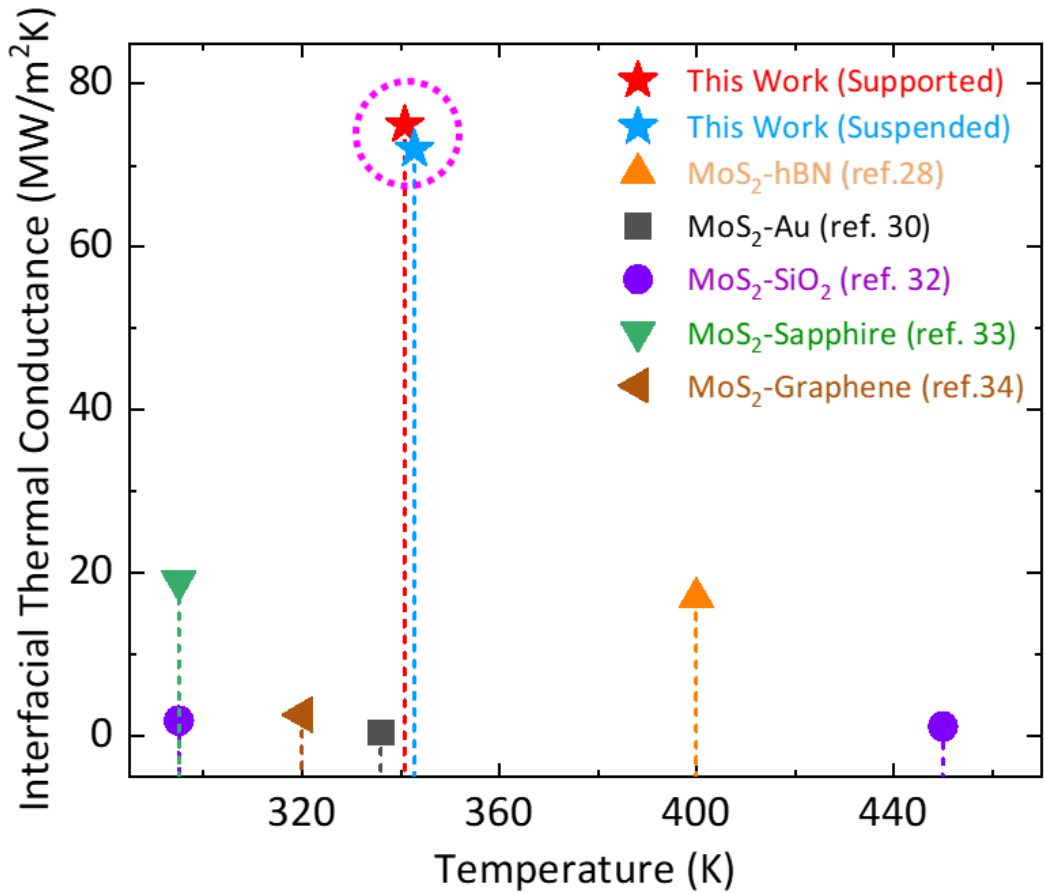

**Figure 5. Benchmark of heat thermal management in hBN/MoS$_2$/hBN van der Waals heterostructure in this work.** Comparison of interfacial thermal conductance in hBN encapsulated MoS$_2$ and other device structures.

**Conclusions**

In conclusion, we have reported the first discovery of the thermal transport properties of hBN encapsulated MoS$_2$ van der Waals heterostructure, and revealed the very high interfacial thermal conductance between hBN and MoS$_2$. By studying Raman spectroscopy under different temperatures and applied laser power, lateral thermal conductivity of hBN and MoS$_2$ and interfacial thermal conductance between different interfaces are discovered. In particular, we found that the interfacial thermal conductance in hBN/MoS$_2$/hBN encapsulated structure is 72 ± 22 MW/m$^2$K in the supported area and 74 ± 25 MW/m$^2$K in the suspended area, which are significantly higher than other structures. The estimated interfacial thermal conductance is consistent with theoretical analysis using MD simulation. These results demonstrate that the hBN encapsulated structure provides efficient heat dissipation pathways for encapsulated 2D material and paves way for building high-performance hBN encapsulated 2D materials nanoelectronics.

## Methods

**Suspended hBN/MoS₂/hBN van der Waals Heterostructure Device Fabrication**

The fabrication of hBN/MoS$_2$/hBN van der Waals heterostructure device starts from exfoliating hBN and MoS$_2$ on PDMS. After selecting ideal hBN and MoS$_2$, bottom hBN, MoS$_2$, top hBN are transferred onto pre-patterned substrates with micro trenches using an all-dry transfer method with careful alignments under microscope [37].

**Raman Scattering and Photoluminence Measurement**

The hBN/MoS$_2$/hBN heterostructure device is measured using a RENISHAW InVia Raman Microscope system with a 514 nm laser. In the temperature dependent Raman measurement, heterostructure samples are heated uniformly from 295K to 495K using a heating stage (Linkam Stage THMS600). The laser power is kept below 100μW during the measurement to reduce additional laser heating. In both temperature dependent measurement and power dependent measurement, the laser is focused on supported and suspended regions.

**Atomic Force Microscopy (AFM) Measurement**

Atomic force microscope (AFM) (Bruker Dimension Icon) is used to measure the thickness of top and bottom hBN layers. The AFM tapping mode in air is used.

**Acknowledgement**: We thank the financial support from Stevens Institute of Technology's startup funding, Stevens Institute of Technology's Bridging Award, and Brookhaven National Laboratory Center for Functional Nanomaterials. The work at Case Western Reserve University is supported by National Science Foundation CAREER Award (Grant ECCS-1454570) and CCSS Award (Grant ECCS-1509721).

**Corresponding Author**
*Baoxing Xu. Email: bx4c@virginia.edu
*Philip Feng. Email: philip.feng@ufl.edu
*Xian Zhang. Email: xzhang4@stevens.edu